\documentclass[5p,times]{elsarticle}

\usepackage[utf8]{inputenc}
\usepackage{listings}
\usepackage{graphicx}
\usepackage{url}
\usepackage{amsmath}
\usepackage{algorithm}
\usepackage{algorithmic}
\usepackage{hyperref}

\newcommand{\genasis}{{\sc GenASiS}}
\newcommand\blfootnote[1]{%
  \begingroup
  \renewcommand\thefootnote{}\footnote{#1}%
  \addtocounter{footnote}{-1}%!TEX encoding = UTF-8 Unicode
  \endgroup
}

\begin{document}
%opening
%\title{Targetting Accelerators for Astrophysics Code \genasis\ with OpenMP Directives on Summit}

\begin{frontmatter}
 
\title{Targeting GPUs with OpenMP Directives on Summit: \\
A Simple and Effective Fortran Experience}

\author[a,c]{Reuben D. Budiardja\corref{author}}
\author[b,c]{Christian Y. Cardall}

\cortext[author] {Corresponding author.\\\textit{E-mail addresses:} reubendb@ornl.gov (R.D. Budiardja), cardallcy@ornl.gov (C.Y. Cardall).}
\address[a]{National Center for Computational Sciences, Oak Ridge National Laboratory, Oak Ridge, TN 37831-6354, USA}
\address[b]{Physics Division, Oak Ridge National Laboratory, Oak Ridge, TN 37831-6354, USA}
\address[c]{Department of Physics and Astronomy, University of Tennessee, Knoxville, TN 37996-1200, USA}

\begin{abstract}

We use OpenMP to target hardware accelerators (GPUs) on Summit, a newly deployed supercomputer at the Oak Ridge Leadership Computing Facility (OLCF), demonstrating simplified access to GPU devices for users of our astrophysics code \genasis\ and useful speedup on a sample fluid dynamics problem.
%At a lower level, we use the capabilities of Fortran 2003 for \texttt{C} interoperability to provide wrappers to the OpenMP device memory runtime library routines (currently available only in $\texttt{C}$).
We modify our workhorse class for data storage to include members and methods that significantly streamline the persistent allocation of and association to GPU memory.
Users offload computational kernels with OpenMP \texttt{target} directives that are rather similar to constructs already familiar from multi-core parallelization.
In this initial example we ask, ``With a given number of Summit nodes, how fast can we compute with and without GPUs?'', and find total wall time speedups of $\sim 12\mathrm{X}$.
We also find reasonable weak scaling up to 8000 GPUs (1334 Summit nodes).
%in `proportional resource tests' that compare runs with a given percentage of nodes' GPUs with runs utilizing instead the same percentage of nodes' CPU cores, and reasonable weak scaling up to 8000 GPUs vs. 56,000 CPU cores ($1333\frac{1}{3}$ Summit nodes). 
We make available the source code from this work at \texttt{https://github.com/GenASiS/GenASiS\_Basics}.
\end{abstract}

\end{frontmatter}

% to avoid \texttt overruns
\sloppy 

\blfootnote{\textbf{Notice of copyright}: This manuscript has been authored by UT-Battelle, LLC under Contract No. DE-AC05-00OR22725 with the U.S. Department of Energy. The United States Government retains and the publisher, by accepting the article for publication, acknowledges that the United States Government retains a non-exclusive, paid-up, irrevocable, worldwide license to publish or reproduce the published form of this manuscript, or allow others to do so, for United States Government purposes. The Department of Energy will provide public access to these results of federally sponsored research in accordance with the DOE Public Access Plan (http://energy.gov/downloads/doe-public-access-plan).}

\section{Introduction}
\label{sec:Introduction}

%\begin{itemize}
% \item Why OpenMP: 
% Allow us to use Fortran features and write in Fortran semantics without having to worry (too much) about how memory are addressed under the hood. Example of these includes: multi-dimensional (3D in our case) arrays, using feature such as pointer remapping. This also allow us to 'fall back' to CPU code.
% 
% \item Recent availability of compiler support
%\end{itemize}

As of version 4.5, OpenMP provides an excellent opportunity to access the extraordinary computational power of the GPUs on machines like Summit at the Oak Ridge Leadership Computing Facility (OLCF).
Most notably, the similarity to existing OpenMP multi- or many-core parallelization coding patterns allows relative ease of porting and minimizes divergence between versions of a code to be run with or without GPUs or other hardware acceleration devices.

That the programmer need not engage the CUDA programming model is particularly useful in Fortran applications.
Given the broad acceptance and entrenched use of OpenMP, it can be expected that the facilities for devices now specified in its standard will receive wide support. 
This is in contrast for instance with CUDA Fortran, a non-standard Fortran extension provided by only two compiler vendors (PGI and IBM XL), excluding many widely-used compilers (e.g. GCC, Intel, and Cray). 
Moreover, OpenACC---an alternative directive-based approach---is currently actively supported only by the PGI compiler, and some ongoing efforts in GCC.
On Summit in particular, excellent support for OpenMP 4.5 already exists with the IBM XL compiler. 
Our experience so far is that Fortran features and semantics such as multidimensional arrays, array sections, and pointer remapping (allowing for instance access to a given memory block as either a 3D or 1D array) are translated just as one would hope and expect.

We have recently begun using OpenMP to adapt our code \genasis\ (\textit{Gen}eral \textit{A}strophysical \textit{Si}mulation \textit{S}ystem) to the exploitation of GPUs on Summit.
Aggressively deploying the features introduced in Fortran 2003 that facilitate an object-oriented approach, \genasis\ is an extensible multiphysics simulation code aimed primarily at astrophysics applications \cite{cardall_2014}.
Early versions have been used to study aspects of the post-bounce core-collapse supernova environment \cite{endeve_2010,endeve_2012,endeve_2013,cardall_2015,budiardja_2015}.
\genasis\ is divided into three primary divisions: \texttt{Basics}, which includes \textit{utilitarian functionality} generically needed by physics simulations on distributed-memory supercomputers \cite{cardall_2015a,cardall_2017}; \texttt{Mathematics}, 
which encompasses implementations of \textit{manifolds} (i.e. meshes), \textit{operations}, and \textit{solvers} \cite{cardall_2018}; and \texttt{Physics}, which comprises \textit{spaces} governed by various theories of spacetime and gravity, specific forms of \textit{stress energy} such as fluids and radiation, and the combination of these into \textit{universes}.
See Figures 1 and 2 of reference \cite{cardall_2015a} for the overall structure of \genasis\, and of \genasis\ \texttt{Basics}, which is the basis of this work. 

In this paper we report our initial explorations, which include implementation of GPU-related infrastructure in \genasis\ \texttt{Basics} and its use in a simple fluid dynamics problem, the \texttt{RiemannProblem} example built only upon \texttt{Basics} classes.
We assume throughout that the reader has some basic familiarity with OpenMP 4.5 device-related constructs.\footnote{For a thorough description of OpenMP \texttt{target} and device-related constructs, the reader is referred to the OpenMP 4.5 specification, reference guide, and examples available at https://www.openmp.org/specifications/.}
In Section~\ref{sec:UsingDevice} we describe the functionality we have added to \genasis\  to streamline the use of GPUs.
In Section~\ref{sec:Porting} we introduce the \texttt{RiemannProblem} example and describe its porting to GPU-aware code.
Section~\ref{sec:Performance} reports the performance gains we obtain on Summit by exploiting its GPU resources.
In Section~\ref{sec:BeyondOpenMP} we show further performance gains that can be achieved when later OpenMP specifications adopt more advanced memory management.
Section~\ref{sec:Conclusion} offers some concluding thoughts on the state of OpenMP implementations vis-\`a-vis GPU usage.
In what follows we use the term `host' to refer to the set of cores assigned to a Message Passing Interface (MPI) process on the CPU, and `device' to refer to the GPU available to the same MPI process. 
%(We always associate one GPU with a single MPI process.)
(In its current version, our code can only use one GPU per MPI process.)

We make available the source code from this work at \texttt{https://github.com/GenASiS/GenASiS\_Basics}.

%\section{Device Memory: Allocation, Association, and Transfer}
%\section{Using a Device in \genasis}
\section{Using an Accelerator Device in \genasis}
\label{sec:UsingDevice}

In many cases (which may be opaque to or unexpected by the programmer), OpenMP directives enable the automatic transfer of data needed by offloaded kernels from the host to the device and back without intervention by the programmer; 
but because large data transfers incur significant overhead, it behooves applications to instead take affirmative control of significant data movement.  
In particular, it is possible to persistently allocate memory on the device for predictively repeated use; to flexibly associate Fortran variables with previously allocated memory locations on the device in a persistent manner; and to command the transfer of data between the host and device.
%When data is present on the device, and host variables are associated with device locations, a computational kernel expressed in terms of those host variables can be executed on the device by enclosing the kernel code between OpenMP directives containing the \texttt{target} keyword.
When data is present on the device, and host variables are associated with device locations, a computational kernel expressed in terms of those host variables can be executed on the device by enclosing the kernel code with the OpenMP \texttt{target} directive and its corresponding \texttt{end target} exit directive.

In this section we distinguish between \textit{lower-level} and \textit{higher-level} functionality in \genasis.
Lower-level functionality consists of the most basic processes implemented as standalone routines.
Higher-level functionality consists of the public methods of a class (using the type-bound procedure facility of the Fortran 2003 standard) for the convenience of programmers writing a \genasis\ application.

\subsection{Higher-level \genasis\ Functionality}
\label{sec:HigherLevel}

The primary way we have made device functionality available to \genasis\ application programmers is by adding members and methods to our workhorse class \texttt{StorageForm}.
This class includes members for both data and metadata; it is used to group together a set of related physical fields on a computational domain.
Code for tasks like I/O, ghost cell exchange, prolongation and restriction on multilevel meshes, and so on are simplified and rendered more generic by use of this class. 

The data member \texttt{Value} of \texttt{StorageForm} is a rank-2 array whose first dimension (rows)---that corresponds to contiguous storage in Fortran---indexes different values of a single physical field at different points in space, with the second dimension (columns) indexing different physical fields.
Metadata members include such things as names and units associated with the several physical fields.

Prior to the addition of device functionality, the only method of \texttt{StorageForm} was an overloaded \texttt{Initialize} procedure.
One mode of initialization entails allocation of the data member \texttt{Value} according to a specified shape; this is a `primary' instance of \texttt{StorageForm}.
A second mode initializes an `overlay' instance of \texttt{StorageForm}, which does not allocate new storage; instead, it points its \texttt{Value} member to that of an existing instance, and includes an integer array member \texttt{iaSelected}\footnote{The prefix \texttt{ia-} in \texttt{iaSelected} is our shorthand for `index array.'}  specifying a subset of physical fields---that is, the indices of selected columns of \texttt{Value} to be regarded as `active' in the new instance. 
(The \texttt{iaSelected} member of a `primary' instance of \texttt{StorageForm} automatically includes all columns of data.)

Adding just a handful of members and methods 
%a member \texttt{D\_Selected} and methods \texttt{AllocateDevice}, \texttt{UpdateDevice}, and \texttt{UpdateHost} 
to our \texttt{StorageForm} class 
provides for the persistent allocation of device memory---and references thereto---in a manner that is simple to use, yet still quite generic and therefore powerful.
A new method \texttt{AllocateDevice()} is used 
%in conjunction with a new \texttt{private} member \texttt{D\_Selected}, an array of \texttt{type(c\_ptr)},
to allocate device memory in a manner that mirrors our usual access to physical fields in primary and overlay instances of \texttt{StorageForm} on the host.
In the case of a primary instance, the method \texttt{AllocateDevice()} (a type-bound procedure of the class) calls the lower-level \texttt{AllocateDevice()} routine (a free-standing wrapper discussed in Section~\ref{sec:LowerLevel}) to allocate a block of memory corresponding to the member \texttt{Value} of \texttt{StorageForm}. 
The sections of this array on the device are persistently associated with the corresponding individual physical fields (columns of the \texttt{Value} member)  on the host by a call to the lower-level \texttt{AssociateHost()} routine (another free-standing wrapper discussed in Section~\ref{sec:LowerLevel}).
In this way, when individual physical fields are addressed inside an OpenMP \texttt{target} region as host variables, the correct device memory locations are referenced.

A call to \texttt{AssociateHost()} for each individual variable is needed following the memory allocation on the device for the following reason. 
Using an unstructured directive
\begin{lstlisting}
!\$OMP target enter data map ( alloc : Value )}  
\end{lstlisting}
to associate the member \texttt{Value} on the host with its copy on the device would not allow us to address individual physical fields (columns of \texttt{Value}).
Although the latter could have been accomplished by instead using the specified columns of \texttt{Value} as the argument to the directive, such as, 
\begin{lstlisting}
do iV = size ( Value, dim = 2 )
  Variable => Value ( :, iV )
  !\$OMP target enter data map ( alloc : Variable )  
end do
\end{lstlisting}
such code may result in memory allocation on the device that is non-contiguous from one column of \texttt{Value} to the next.
By explicitly allocating memory on the device ourselves to the full size of \texttt{Value} (rows and columns), we ensure a contiguous memory allocation that facilitates flexibility in transferring data either in bulk or as individual columns, and in performing operations across multiple columns of \texttt{Value} that may be much more efficient with coalesced memory.

In the case of an overlay instance of \texttt{StorageForm}, a call to the \texttt{AllocateDevice()} method is neither needed nor appropriate, because persistent association of individual physical fields must be established by a call to the \texttt{AllocateDevice()} method of the primary instance.

Finally, the methods \texttt{UpdateDevice()} and \texttt{UpdateHost()} (which call the lower-level routines of corresponding name discussed in Section~\ref{sec:LowerLevel}) allow all the selected fields of an instance of \texttt{StorageForm} to be transferred to or from the device en masse with a single call.
Alternatively, individual fields can be specified for transfer.

\subsection{Lower-level \genasis\ functionality}
\label{sec:LowerLevel}

In the previous section we explained why we need to use OpenMP runtime library routines for device memory management---such as the persistent allocation of memory on the device and for association of a host variable with previously allocated device memory---instead of \texttt{target enter data} directives.
Because these routines are only provided in C, we use the \texttt{C} interoperability capabilities of Fortran 2003 to write simple Fortran wrappers to make this functionality available as standalone routines \genasis.

In our code, the line
\begin{lstlisting}
call AllocateDevice ( Value, D_Value )
\end{lstlisting}
allocates a memory region on the device corresponding to the size of the Fortran array \texttt{Value} previously allocated on the host, and sets the variable \texttt{D\_Value} of \texttt{type(c\_ptr)} to the memory location on the device.
\footnote{As might be guessed from the paired variable names \texttt{Value} and \texttt{D\_Value}, our default approach is to mirror variables allocated in host memory on the device, using the prefix \texttt{D\_} to denote the device copy.}
Similarly, the command  
\begin{lstlisting}
call AssociateHost ( D_Value,  Value )
\end{lstlisting}
causes references to a host variable \texttt{Value} appearing in an appropriate OpenMP directive to be interpreted as referring to the device memory location pointed to by \texttt{D\_Value}.
We have also provided the corresponding Fortran routines \texttt{DeallocateDevice()} (which frees memory on the device pointed to by an argument \texttt{D\_Value}) and \texttt{DisassociateHost()} (which frees a host variable argument \texttt{Value} from its association with a memory location on the device).
Under the hood, these wrappers call the OpenMP \texttt{C} functions \texttt{omp\_target\_alloc()}, \texttt{omp\_target\_associate\_ptr()}, \texttt{omp\_target\_free()}, and \texttt{omp\_target\_disassociate\_ptr()} respectively, making use of the Fortran 2003 \texttt{bind} attribute.
In each case the function \texttt{omp\_get\_default\_device()} is used to identify the device, which is all that is needed with our convention of only one device (i.e. GPU) per MPI process.

We also provide standalone Fortran wrappers in \genasis\ to transfer data to and from the device, with the commands
\begin{lstlisting}
call UpdateDevice ( Value, D_Value )
\end{lstlisting}
and
\begin{lstlisting}
call UpdateHost ( D_Value, Value )
\end{lstlisting}
performing transfers from host to device and from device to host respectively. Similarly, under the hood these wrappers call the OpenMP \texttt{C} function \texttt{omp\_target\_memcpy()}.

Although in implementing these lower-level routines we have used OpenMP runtime library routines under the hood, in principle they constitute a fa\c{c}ade allowing implementation using other backends, such as the Nvidia CUDA or AMD ROCm libraries. 

Normally, we expect that \genasis\ application programmers will not need to use these lower-level routines, and therefore never have to deal with device pointers.
Instead, we expect application programmers to use the methods of \texttt{StorageForm} for allocation of and association to device memory, and for transfers between host and device, as discussed in Section \ref{sec:HigherLevel}.

\subsection{Offloading Computational Kernels in \genasis} 
\label{sec:Offloading}

\begin{lstlisting}[float,frame=tb,numbers=left,numbersep=5pt,xleftmargin=10pt,label=lst:AddKernel,caption={An example computational kernel.}]
subroutine AddKernel ( A, B, C )

  real ( KDR ), dimension ( : ), intent ( in ) :: A, B
  real ( KDR ), dimension ( : ), intent ( out ) :: C
  
  integer ( KDI ) :: i
  
  !\$OMP target teams distribute parallel do schedule ( static, 1 )
  do i = 1, size ( C )
    C ( i )  =  A ( i )  +  B ( i )
  end do
  !\$OMP end target teams distribute parallel do

end subroutine AddKernel
\end{lstlisting}

In contrast to the functionality added to our workhorse \texttt{StorageForm} class as discussed in Section~\ref{sec:HigherLevel}---which shields application programmers from direct references to OpenMP---the offloading of computational kernels to the device requires OpenMP directives.
This is to be expected given the inherent nature of computational kernels, defined as straightforward segments of code implementing repetitive basic operations, through which significant amounts of data are fed.

In line with common practice and in part for traditional purposes of modularity, our habit is to sequester computational kernels in dedicated subroutines, to which only a few changes are needed to achieve offloading.
In \genasis\ the sequestration of kernels in subroutines also serves---via arrays provided in argument lists---the additional purpose of exposing fundamental data types to both the programmer and the compiler.
That is, instead of confronting a human or algorithmic parser with data buried in more complicated structures, such as array sections of the rank-2 pointer member \texttt{Value} of our \texttt{StorageForm} class, programmers and compilers see straightforward arrays of real variables.

Suppose for example we have an instance \texttt{Fields} of class \texttt{StorageForm} whose \texttt{Value} member contains three columns of data corresponding to three physical fields.
Adding the line 
\begin{lstlisting}
call Fields % AllocateDevice ( )
\end{lstlisting} 
to the program initialization allocates a block of contiguous device memory comprising the three fields, and associates the individual fields on the host with the appropriate addresses within the memory block on the device.
Suppose further that we wish to set the third field to the sum of the first two fields by executing a routine \texttt{AddKernel()} on the device.
The data is originally on the host, and is again needed on the host after the operation.
The lines
\begin{lstlisting}
call Fields % UpdateDevice ( )
call AddKernel &
       ( Fields % Value ( :, 1 ), &
         Fields % Value ( :, 2 ), &
         Fields % Value ( :, 3 ) )
call Fields % UpdateHost ( )
\end{lstlisting}
accomplish this task.

The computational kernel called here, shown in Listing~\ref{lst:AddKernel}, has only one element that distinguishes it from a kernel not intended for offloading to the device.\footnote{The parameters \texttt{KDR} in lines~3 and 5 and \texttt{KDI} in line~7 specify the Fortran \texttt{kind} of \texttt{real} and \texttt{integer} variable respectively.}

The kernel loop in lines~8-12 is accessorized with an OpenMP \texttt{target} directive rather than a similar construct
\begin{lstlisting}
!\$OMP parallel do schedule ( static, 1 )
\end{lstlisting}
that would be used in a traditional CPU multithreading context. 
The \texttt{target} directive in line 8 tells OpenMP to offload the computational kernel (line 9-11) to the device.
Because the host variables \texttt{A}, \texttt{B}, and \texttt{C} have been previously associated with the device memory location inside the method \texttt{Fields \% AllocateDevice ( )}, the OpenMP runtime uses the corresponding device locations for those variables, avoiding any implicit data transfer.
The \texttt{teams distribute} directive in the same line creates and distributes work to a league of teams of threads on the device.
The manner in which the teams of threads are mapped to the device hardware is implementation-dependent.

It may be asked why the transfer of data to and from the device is not included in the kernel itself---perhaps even left to be done implicitly and automatically by the compiler---and the answer is important to the simple and effective porting of sophisticated codes.
Because of the nontrivial overhead involved in data transfer to and from the device, it is highly desirable to keep data on the device for as long as possible. 
The above simple example involves only a single kernel, but realistic cases may involve calls to many separate kernels between necessary updates to the host.
The persistent allocation and association of device memory allows 
data to be left on the device for extended operations involving multiple separate kernel calls without disrupting the existing structure of thoroughly modularized codes.

% \begin{itemize}
%  \item \sout{Keeping data on GPU} (FIRST PARAGRAPH OF Section~2; Section~2.3)
%  \item \sout{Data association with device pointer for explicit control of data movement} (FIRST PARAGRAPH OF Section~2)
% \end{itemize}

% \sout{(Once we worked out what we needed to do,) We follow a fairly simple prescription in porting our code to use GPU using OpenMP. 
% In this subsection we start by outlining our prescription, followed by a more detailed elaboration for each step of the prescription.} 

% \sout{Our porting efforts can be summarized by the following steps:}
%\begin{enumerate}
%  \item \sout{Allocate the necessary memory space in GPU at the beginning of the program, mirroring the memory allocation on the CPU.} (SECS.~2.1 and 2.2) 
%  \item \sout{Associate array variables on the host with the pointers storing the corresponding addresses for those variables on the GPU to avoid automatic data-movement.} (FIRST PARAGRAPH OF Section~2; Section~2.1; Section~2.3)
%  \item \sout{Accesorize loops with OpeMP \texttt{target} directive for execution on the GPU.} (FIRST PARAGRAPH OF Section~2; Section~2.3)
%  \item \sout{Remove association of the array variables with the device pointer.} (Section~2.1; Section~2.3)
% \end{enumerate}

\section{Porting a Fluid Dynamics Application: \texttt{RiemannProblem}}
\label{sec:Porting}
\begin{figure}
\centering
\includegraphics[width=0.22\textwidth]{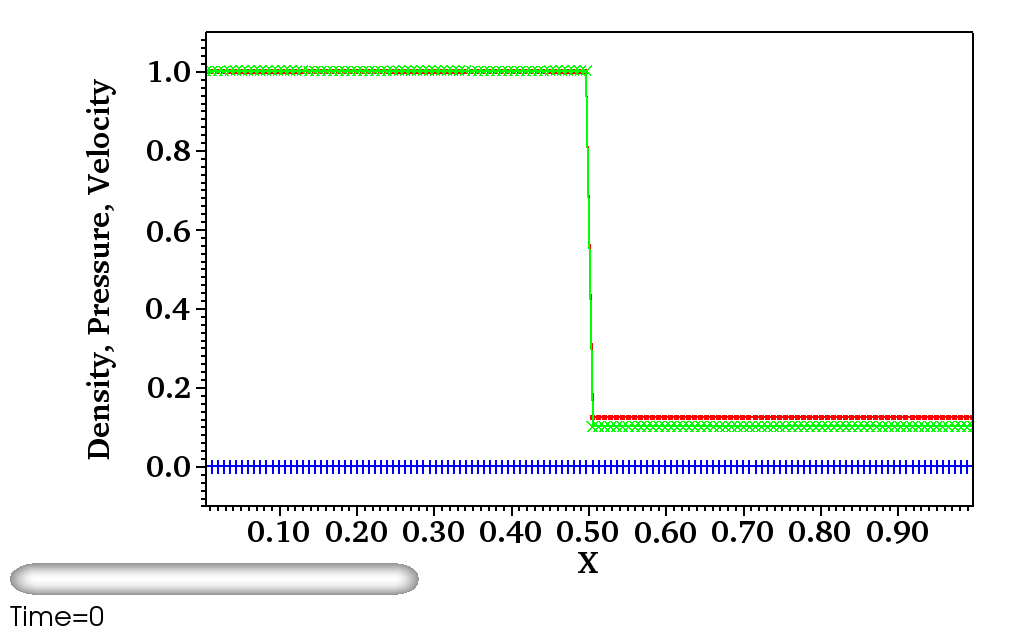}
\includegraphics[width=0.22\textwidth]{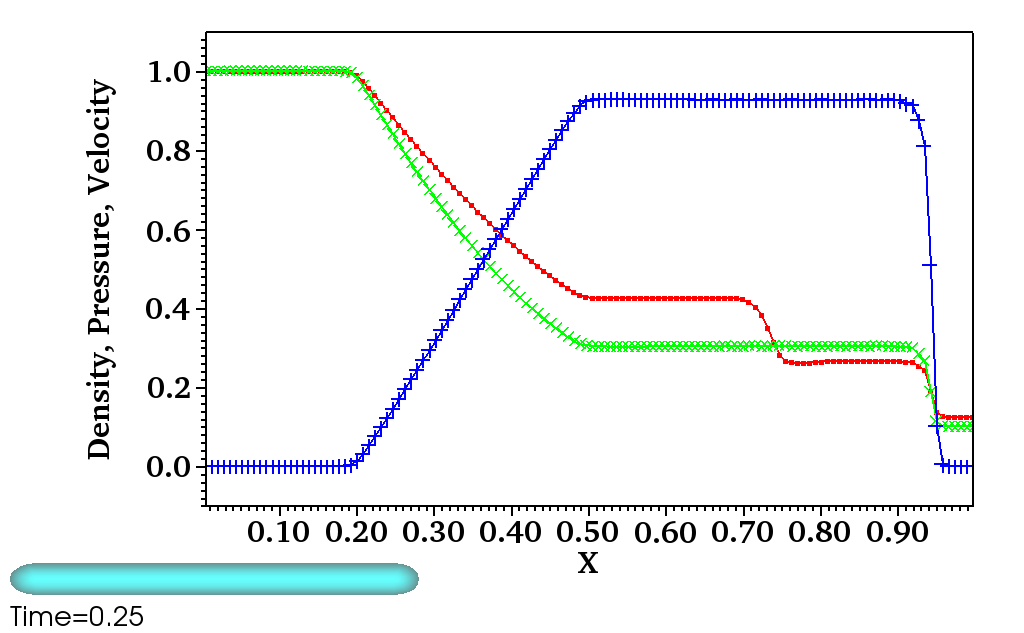}\\
\vspace{1cm}
\includegraphics[width=0.22\textwidth]{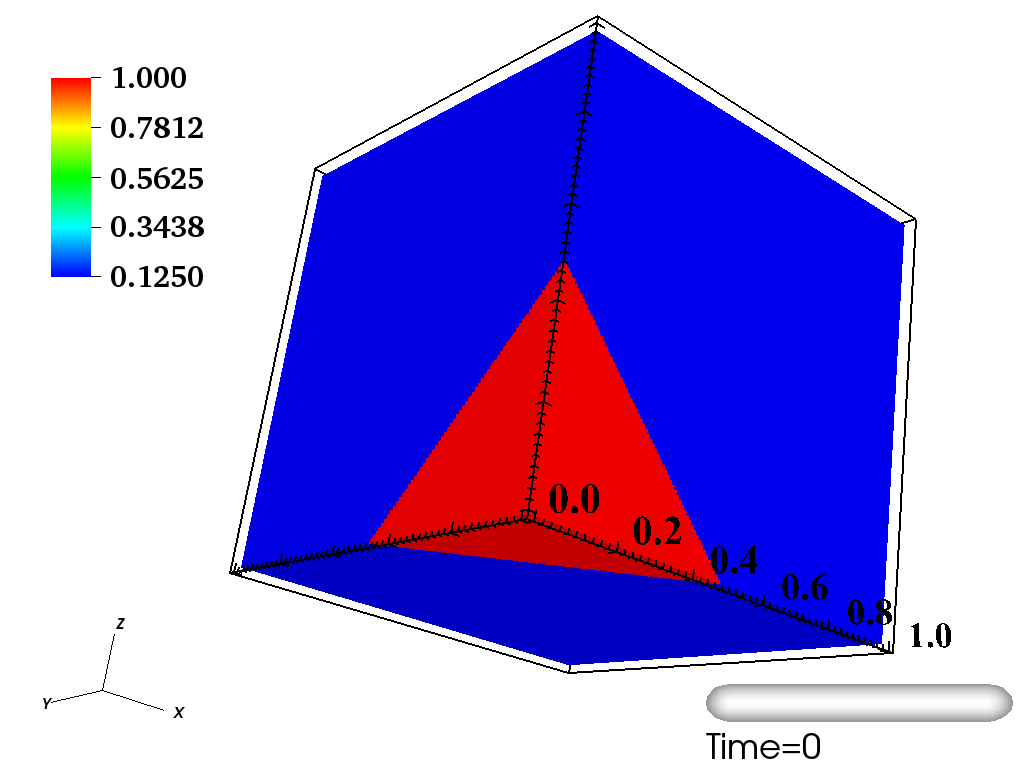}
\includegraphics[width=0.22\textwidth]{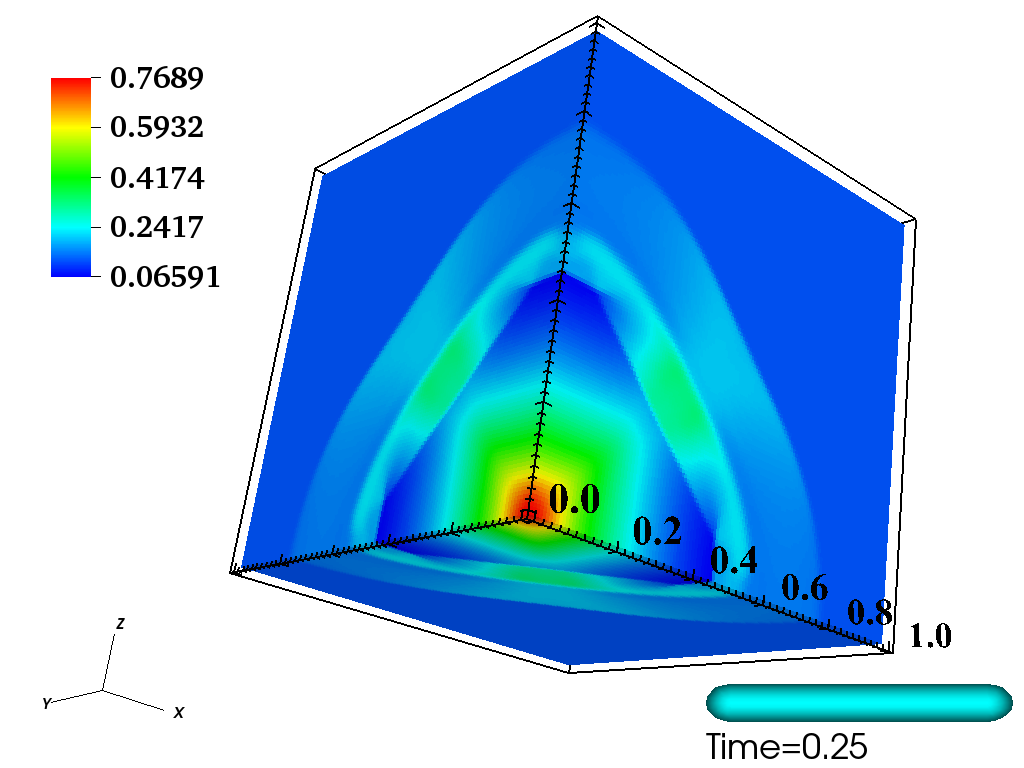}
\caption{The initial (left) and final (right) density in 1D (upper) and 3D (lower) versions of \texttt{RiemannProblem} evolved to time $t=0.25$ at a resolution of $128$ cells in each dimension. 
In the 1D case, the density (red) is plotted along with pressure (green) and velocity (blue).}%, with one symbol for every four cells in each case.}
\label{fig:RiemannProblem}
\end{figure}

As a concrete working example of targeting GPUs with OpenMP directives, we use our implementation to solve a \texttt{RiemannProblem} using \genasis\ \texttt{Basics}.
\texttt{RiemannProblem} is an extension of the classic 1D Sod shock tube---a standard computational fluid dynamics test problem---to 2D and 3D. Figure~\ref{fig:RiemannProblem} shows the initial and final state of 1D and 3D versions of \texttt{RiemannProblem} after being evolved to time $t=0.25$ at a resolution of $128$ cells in each dimension. 

Pseudocode outlining the solution in \texttt{RiemannProblem} (and other simple fluid dynamics examples distributed with \genasis\ \texttt{Basics}) is displayed in Algorithm~\ref{alg:Evolve}.
\begin{algorithm}
  \begin{algorithmic}[1]
    \STATE {\em Call: } Initialize ( )
    \STATE {\em Call: } GhostExchange ( )
    \STATE {\em Set: } Time = StartTime
    \WHILE{Time $<$ FinishTime}
    \STATE {\em Call: } ComputeTimeStep ( ) $\rightarrow$ TimeStep
    \STATE {\em Set: } FluidOld = FluidCurrent
    \STATE
    \STATE {\em Call: } ComputeDifferences ( )
    \STATE {\em Call: } ComputeReconstruction ( )
    \STATE {\em Call: } ComputeFluxes ( )
    \STATE {\em Call: } ComputeUpdate ( TimeStep ) $\rightarrow$ FluidUpdate
    \STATE {\em Set: } FluidCurrent = FluidOld + FluidUpdate
    \STATE
    \STATE {\em Call: } GhostExchange ( )
    \STATE
    \STATE {\em Call: } ComputeDifferences ( )
    \STATE {\em Call: } ComputeReconstruction ( )
    \STATE {\em Call: } ComputeFluxes ( )
    \STATE {\em Call: } ComputeUpdate ( TimeStep )
    \STATE {\em Set: } FluidCurrent = 0.5 * (FluidOld + FluidCurrent + FluidUpdate)
    \STATE
    \STATE {\em Call: } GhostExchange ( )
    \ENDWHILE
    \caption{Fluid Evolution}
    \label{alg:Evolve}
  \end{algorithmic}
\end{algorithm}
Each iteration of the main loop enclosed by lines~4 and 23 is a second-order Runge-Kutta time step consisting of two substeps, each of which consists of several computational kernels (lines~8-12 and 16-20).
Because the emphasis in the \texttt{Basics} examples as originally distributed \cite{cardall_2015a,cardall_2017} was to demonstrate use of the utilitarian functionality provided by that division of the code, the kernels were written in the most simple-minded and compact manner possible, with essentially no regard for performance or even traditional OpenMP multi-core threading.
Eschewing explicit loops and indexing, Fortran array syntax---whole-array operations, the \texttt{where} construct, the \texttt{cshift()} operation, and the like---had been used throughout.
Therefore our first task, prior to beginning the porting required for exploitation of GPU devices, was to rewrite these kernels in terms of \texttt{do} loops over explicitly indexed arrays, accessorized with OpenMP directives implementing multi-core threading.
This then constituted a code that provided a basis for a speedup comparison, with kernels readily adaptable to offloading via modification of the familiar OpenMP \texttt{parallel do} construct. 

Given the infrastructure and approach to accelerator devices described in Section~\ref{sec:UsingDevice}, porting Algorithm~\ref{alg:Evolve} to its device-enabled counterpart in Algorithm~\ref{alg:EvolveAccelerator} was both straightforward and effective.
\begin{algorithm}
  \begin{algorithmic}[1]
    \STATE {\textsc{Host:} \em Call: } Initialize ( ) 
    \STATE {\em Call: } GhostExchange ( )
    \STATE {\textsc{Host:} \em Set: } Time = StartTime
    \WHILE{Time $<$ FinishTime}
    \STATE {\textsc{Host:} \em Call: } ComputeTimeStep ( ) $\rightarrow$ TimeStep
    \STATE {\textsc{Transfer: } \em Call: } FluidCurrent \% UpdateDevice ( )
    \STATE {\textsc{Device:} \em Set: } FluidOld = FluidCurrent
    \STATE
    \STATE {\textsc{Device:} \em Call: } ComputeDifferences ( )
    \STATE {\textsc{Device:} \em Call: } ComputeReconstruction ( )
    \STATE {\textsc{Device:} \em Call: } ComputeFluxes ( )
    \STATE {\textsc{Device:} \em Call: } ComputeUpdate ( TimeStep ) $\rightarrow$ FluidUpdate
    \STATE {\textsc{Device:} \em Set: } FluidCurrent = FluidOld + FluidUpdate
    \STATE
    \STATE {\textsc{Transfer: } \em Call: } FluidCurrent \% UpdateHost ( )
    \STATE {\textsc{Host:} \em Call: } GhostExchange ( )
    \STATE {\textsc{Transfer: } \em Call: } FluidCurrent \% UpdateDevice ( )
    \STATE
    \STATE {\textsc{Device:} \em Call: } ComputeDifferences ( )
    \STATE {\textsc{Device:} \em Call: } ComputeReconstruction ( )
    \STATE {\textsc{Device:} \em Call: } ComputeFluxes ( )
    \STATE {\textsc{Device:} \em Call: } ComputeUpdate ( TimeStep )
    \STATE {\textsc{Device:} \em Set:} FluidCurrent = 0.5 * (FluidOld + FluidCurrent + FluidUpdate)
    \STATE
    \STATE {\textsc{Transfer: } \em Call: } FluidCurrent \% UpdateHost ( )
    \STATE {\em Call: } GhostExchange ( )
    \ENDWHILE
    \caption{Fluid Evolution with Accelerator}
    \label{alg:EvolveAccelerator}
  \end{algorithmic}
\end{algorithm}
Algorithm~\ref{alg:EvolveAccelerator} differs from Algorithm~\ref{alg:Evolve} in its notation of whether steps are executed on the \textsc{Host} or \textsc{Device}, but otherwise matches almost line for line in this high-level perspective, the only additions being the \textsc{Transfer} operations explicitly commanded in lines~6, 15, 17, and 25.
In the present state of our code these transfers are necessary to allow the host to exchange ghost data between MPI processes responsible for separate regions of the spatial domain. 
(Future exploitation of CUDA-aware MPI libraries with hardware-enabled GPUDirect technologies should allow this ghost exchange to proceed directly between devices.)
  
We emphasize that the basic structure of the code has been preserved, and that (with the exception of \texttt{FluidCurrent}) the data utilized by the several kernels is generated on the device and remains on the device throughout each of the Runge-Kutta substeps in lines~9-13 and 19-23 of Algorithm~\ref{alg:EvolveAccelerator}.
Instances of our class \texttt{StorageForm} for \texttt{FluidOld}, \texttt{Differences}, \texttt{Reconstruction}, \texttt{Fluxes} and \texttt{FluidUpdate} are initialized as part of line~1 in both Algorithm~\ref{alg:Evolve} and Algorithm~\ref{alg:EvolveAccelerator}; and in the latter case, addition of a call to the \texttt{AllocateDevice()} method of each of these instances to enable use of the device is rather trivial from a programming perspective.
Without such use of designated persistent device memory and controlled transfers, OpenMP by default ensures that data would---with blissful ignorance on the part of the programmer, but at the cost of debilitating wall time overhead---be automatically shipped back and forth between host and device between each of the several kernel calls.
%And the alternative of combining existing kernels into a monolithic procedure to minimize such automated transfers is repellent both as a programming task and as destructive of modularity, thereby impeding the straightforward implementation of algorithmic variations.
% An alternative would be to combine existing kernels into a monolithic subroutine, and eliminate unwanted transfers via an \texttt{!\$OMP target enter data} directive and appropriate use of \texttt{map} clauses in the \texttt{target} directives enclosing the several loops.
% We find this repellent both as a programming task and as destructive of modularity, thereby impeding straightforward implementation of algorithmic variations.

%FIXME: Should we just remove this sentence since this was described in Sec. 2.2?
%(Achieving this sort of control via OpenMP \texttt{target enter data copyin} and \texttt{target exit data copyout} directives would be inconsistent with the spirit of our higher-level \texttt{StorageForm} functionality, where we have provided \texttt{UpdateHost()} and \texttt{UpdateDevice()} methods.) 

It is worth noting a degree of flexibility available in the mapping of host program variables to device memory in computational kernels.
As discussed in Section~\ref{sec:HigherLevel}, we store the values of a physical field corresponding to different cells of a discretized computational domain in a column of a rank-2 Fortran array, the \texttt{Value} member of class \texttt{StorageForm}.
However in our fluid dynamics problem (and as is of course the case in general), some of the kernels require knowledge of spatial relationships in the data, in particular, nearest-neighbor (or wider) stencils.
In this \texttt{Basics} example problem a single-level rectangular mesh is employed, so that stencil relationships can be represented through appropriate indexing of rank-3 arrays embodying discretized 3D position space.
Such 3D arrays can be obtained from a column of data with the Fortran pointer remapping facility.
For instance, local rank-3 pointer variables \texttt{V} and \texttt{dV} declared as
\begin{lstlisting}
real ( KDR ), dimension ( :, :, : ), pointer :: V, dV
\end{lstlisting}
can be used to remap (say) the \texttt{iV}-th data column of instances \texttt{FluidCurrent} and \texttt{Differences} of class \texttt{StorageForm}: 
\begin{lstlisting}
 V ( -1:nX+2, -1:nY+2, -1:nZ+2 ) &
   => FluidCurrent % Value ( :, iV )
dV ( -1:nX+2, -1:nY+2, -1:nZ+2 ) &
   =>  Differences % Value ( :, iV )
\end{lstlisting} 
where \texttt{nX}, \texttt{nY}, \texttt{nZ} are the numbers of `proper' or active cells in each of the three dimensions of a subdomain assigned to an MPI task, and allowance is made for two layers of ghost cells. 
Now the rank-3 arrays \texttt{V} and \texttt{dV} can be sent to a kernel \texttt{ComputeDifference\_X} shown in Listing~\ref{lst:ComputeDifference}.
\begin{lstlisting}[float,frame=tb,numbers=left,numbersep=5pt,xleftmargin=10pt,label=lst:ComputeDifference,caption={A kernel for computing a nearest-neighbor difference.}]
subroutine ComputeDifference_X ( V, dV )

  real ( KDR ), dimension ( -1:, -1:, -1: ), &
    intent ( in ) :: &
      V
  real ( KDR ), dimension ( -1:, -1:, -1: ), &
    intent ( out ) :: &
      dV
  
  integer ( KDI ) :: i, j, k
  
  !\$OMP target teams distribute parallel do collapse ( 3 ) schedule ( static, 1 )
  do k = 1, nZ
    do j = 1, nY
      do i = 0, nX + 2
        dV ( i, j, k ) &
          =  V ( i, j, k )  -  V ( i - 1, j, k )
      end do
    end do
  end do
  !\$OMP end target teams distribute parallel do

end subroutine ComputeDifferences_X
\end{lstlisting}
Even though the associations between the host variables and device memory locations were originally set in connection with rank-1 arrays (single columns of a \texttt{Value} data array), the references in lines 16 and 17 to rank-3 aliases of these data columns work just as hoped and expected. 
 
%\begin{itemize}
 %\item Introduce Riemann Problem
 %\item Replace any Fortran-style array syntax such as array assignment, \texttt{where} construct, and \texttt{cshift()} with regular \texttt{do-loop} construct. (DISCUSSION OF RIEMANN PROBLEM?)
 %\item Paragraph about pointer remapping and kernel
 %\item 
%FIXME: Talk about compiler and Flags 
%  \item \sout{General code overview} (THIRD PARAGRAPH OF Section 1)
%\end{itemize}

To compile our code, we use IBM XL Fortran, version 16.1.1 with \texttt{-qsmp=omp -qoffload -Ofast} flags. The invocation \texttt{xlf2008\_r} is used to compile Fortran 2008 source files with thread-safe compilation. The MPI library is provided by IBM Spectrum MPI.

\section{Performance Results}
\label{sec:Performance}

%FIXME: Discuss two possible mode of running: GPU sharing with multi-process server (MPS), vs exclusive

The test results described here were performed on Summit, a newly deployed supercomputer at the Oak Ridge Leadership Computing Facility (OLCF).\footnote{https://www.olcf.ornl.gov/for-users/system-user-guides/summit/} 
A compute node on Summit is an IBM Power System AC922, which consists of two IBM Power9 CPUs and six NVIDIA Volta V100 GPUs. 
Each Power9 CPU has 22 physical cores, one of which is reserved for the operating system tasks, leaving 42 physical cores on a Summit compute node available for running a user's application.
Three GPUs and a Power9 CPU are interconnected with NVLINK. 
The two Power9 CPUs are connected by an X-Bus.
Summit nodes are interconnected in a non-blocking fat tree topology with a dual-rail EDR InfiniBand network.

Summit's job launching system, \texttt{jsrun}, allows users to create logical abstractions of Summit nodes with the concept of a \textit{resource set}. 
%For example, a Summit compute node may be logically abstracted to appear as six logical nodes to an application by creating six resource sets per node where where each resource set has seven CPU cores and one GPU. 
%Here we utilize this functionality to set up `proportional resource tests' in which we compare runs with using either the GPUs or the CPU cores in the resource sets. 
Here we create six resource sets per node, where each resource set has seven CPU cores (i.e. 1/6 of the 42 cores per node) and one GPU (i.e. 1/6 of the six GPUs per node).
Each resource set is assigned to one MPI process.
This allows for meaningful speedup measurements in `proportional resource tests' that compare the performance of runs using \textit{either} the GPUs \textit{or} the CPU cores in the resource sets.
In other words, we seek to answer the question ``With a given number of Summit nodes, how fast can we compute with and without GPUs?''
%\subsection{Exclusive GPU Mode}

%For the test runs here, we use the GPU in \textit{exclusive} mode by assigning one GPU for each MPI process.
In the test runs reported here, we use the GPU in \textit{exclusive} mode by assigning one GPU to each MPI process.
%To fully utilize Summit compute node, we set up 6 resource sets (RS) per compute node, where each resource set has 1 GPU and 7 CPUs, totaling to 6 GPUs and 42 CPUs, the full Summit compute node, being utilized by the application.
%To fully utilize the Summit compute nodes, we set up 6 resource sets per compute node, where each resource set has 1 GPU and 7 CPUs, totaling to 6 GPUs and 42 CPUs---the full Summit compute node.
%Within a resource set previously described, we runs one MPI process with the possibility of running up to seven OpenMP threads.
In each resource set (described in the preceding paragraph) we run one MPI process with either the GPU or up to seven OpenMP threads on the seven CPU cores assigned to the set.
%As described in \$ \ref{subsec:porting}, the computational kernels in our code can run either with OpenMP threads or OpenMP \texttt{target} directive for GPU, but not both simultaneously.
As alluded to in Section~\ref{sec:Porting}, the computational kernels in our code can run either with OpenMP threads on the CPU or with OpenMP \texttt{target} directives for the GPU, but not both simultaneously.
The speedup results we report below compare these two modes of operation.

%For the performance and scaling tests, we assign $256^3$ cells per MPI process for 3D \texttt{RiemannProblem} to ensure that the problem size is large enough that each MPI process, either when utilizing multiple OpenMP threads or GPU, is not computationally starved.
For the performance and scaling tests, we assign $256^3$ cells per MPI process for the 3D \texttt{RiemannProblem} to ensure that the problem size is large enough that each MPI process, whether utilizing the GPU or multiple OpenMP threads on the CPU, is not computationally starved.
%For a sense of relative comparisons, in our previous production simulations reported in \cite{cardall_2015} \cite{budiardja_2015}, we assigned $128^3$ cells per MPI process to achieve reasonable time to solution per simulation.
For a sense of comparison, in previous production simulations \cite{cardall_2015,budiardja_2015} we assigned $128^3$ cells per MPI process to achieve reasonable time to solution per simulation.

\begin{figure}
\centering
\includegraphics[width=0.45\textwidth]{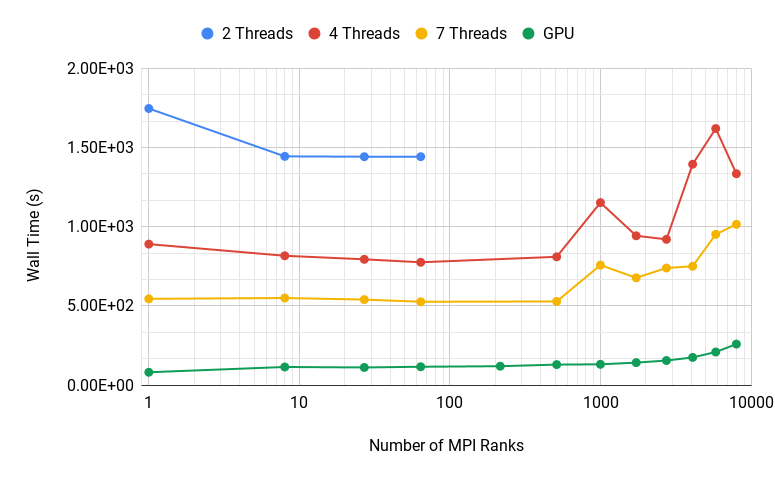}
\caption{Weak scaling of the 3D \texttt{RiemannProblem} with $256^3$ cells assigned to each MPI process.
Wall time includes data transfers between host memory and GPU memory in the GPU version of the code.
Colors correspond to OpenMP thread counts 2 (blue), 4 (red), and 7 (yellow) on the CPU, while the GPU version is shown in green.
Each MPI process is assigned one GPU.
With 8 MPI processes, the speedup from thread counts 2 to 4 to 7 to GPU are $1.76\mathrm{X}$, $1.48\mathrm{X}$, $4.8\mathrm{X}$ respectively.
The speedup from 7 CPU threads to the GPU ranges from $3.92\mathrm{X}$ to $6.71\mathrm{X}$ in these runs.}
\label{fig:3D_WeakScaling}
\end{figure}

%Figure \ref{fig:3D_WeakScaling} shows the weak scaling of 3D \texttt{RiemannProblem} for both versions of the code with OpenMP threads on the CPU and OpenMP \texttt{target} directive on the GPU.
Figure~\ref{fig:3D_WeakScaling} shows the weak scaling of the 3D \texttt{RiemannProblem} for both versions of the code: with OpenMP threads on the CPU, and OpenMP \texttt{target} directives on the GPU.
%As shown on the figure, near-ideal speedup is observed with increasing CPU thread counts from 2, to 4, to 7.
%From 7 CPU threads to GPU, approximately another factor of 4 speedup is observed.
The figure demonstrates near-ideal speedup with increasing CPU thread counts, from 2 to 4 to 7;
and from 7 CPU threads to the GPU, approximately another factor of 6 speedup is observed.
This significant speedup is achieved even when the extra costs of data transfers between the host memory and GPU high-bandwidth memory are included.

\begin{figure*}
\centering
\includegraphics[width=0.96\textwidth]{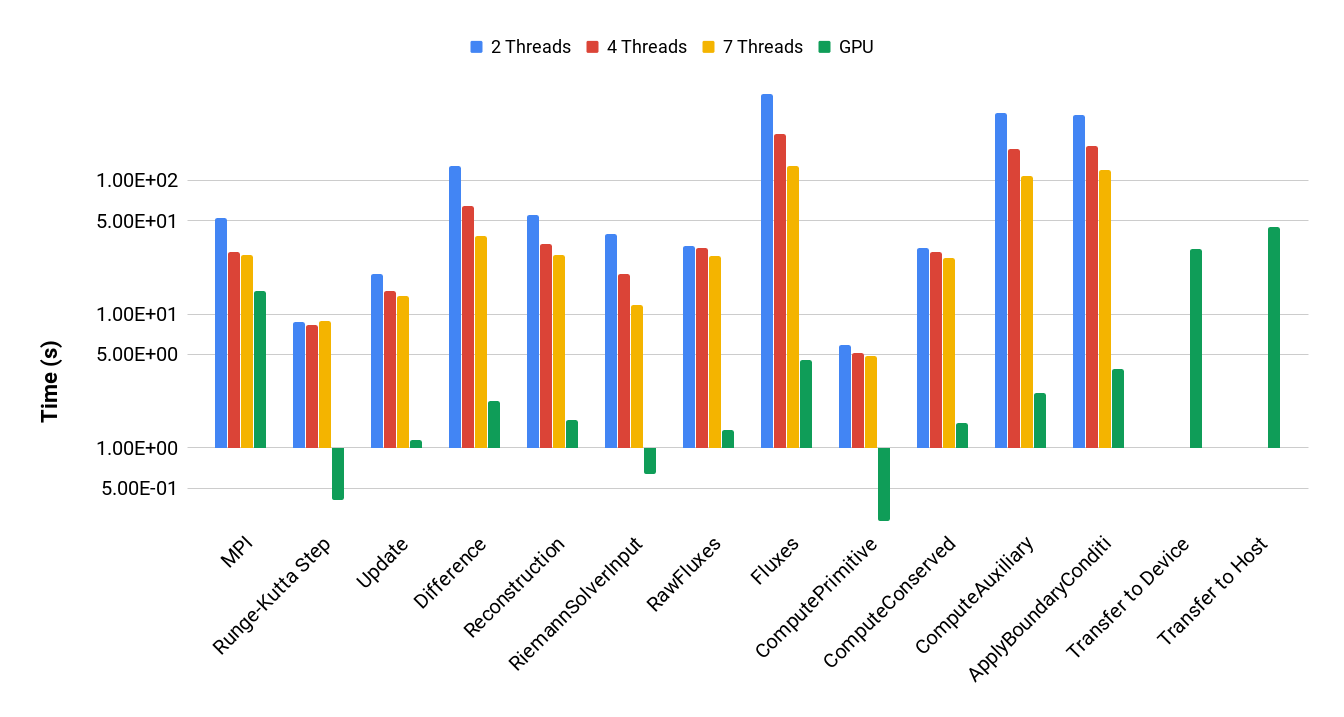}
\caption{%Timings for the different computational kernels, MPI communication, and data transfers to and from GPU in \texttt{RiemannProblem}.
Timings for the MPI communication, computational kernels, and data transfers to and from the GPU in a 3D \texttt{RiemannProblem}.
The blue, red, and yellow timings are for 2, 4, and 7 CPU threads, while the green bars are for the GPU.
All timings are the mean values of the eight MPI processes used in this particular run.}
\label{fig:kernel_timings}
\end{figure*}

To better understand the computational costs of the different portions of our code, we heavily accessorize computational kernels and logical portions of the code with timers from our \texttt{TimerForm} class \cite{cardall_2017}.
%Figure \ref{fig:kernel_dist} shows the timings for computational kernels making up the 3D \texttt{RiemannProblem}, MPI communication, and data transfers to and from GPU for the GPU version of the code.
Figure~\ref{fig:kernel_timings} shows the timings for the computational kernels, MPI communication, and---for runs utilizing GPUs---data transfers to and from the GPUs.
%On the figure we observe that in almost all computational kernels the GPU version outperforms the CPU version with 7 OpenMP threads.
The figure shows that in almost all computational kernels the GPU runs outperform the CPU runs.
%Because the MPI communication portions of the code is single threaded and only runs on the host, we do not expect to see any difference in timing there.
We do not expect significant differences in the timing of the MPI communication portion of the code, which is single-threaded and only runs on the host.
We therefore attribute the timing differences seen in Figure \ref{fig:kernel_timings} for the MPI portion to system noise and general variability. 
(Similar effects of system noise and variability can be seen on the weak scaling plot on Figure \ref{fig:3D_WeakScaling} for runs with 1000 and more MPI processes.)

\begin{figure}
\centering
\includegraphics[width=0.46\textwidth]{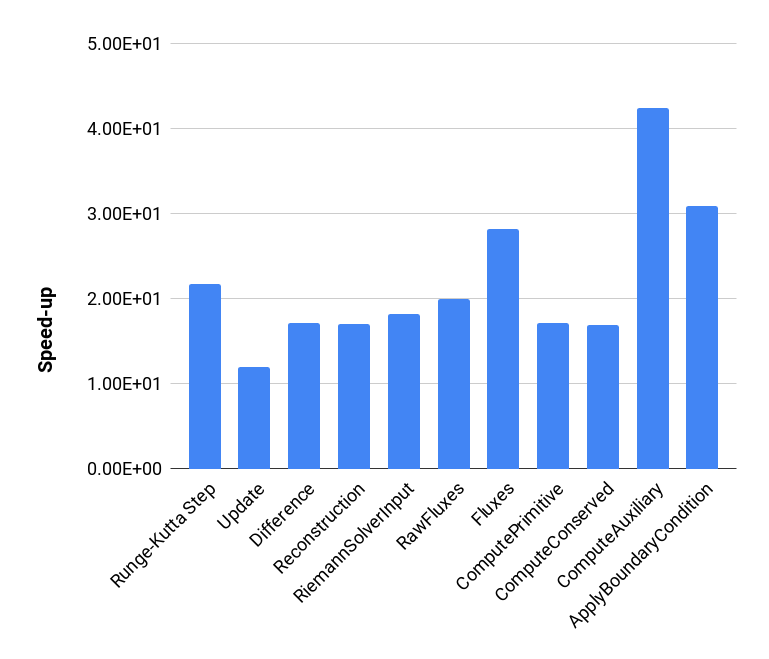}
\caption{%Speed-up for the different computational kernels in \texttt{RiemannProblem} due to GPU from seven CPU threads.
Speedup from the GPU relative to seven CPU threads for the different computational kernels in a 3D \texttt{RiemannProblem}.
Multiple kernels achieve significant speedup, while the kernel labeled \texttt{RiemannSolverInput} gets very little speedup.} 
\label{fig:kernel_speedup}
\end{figure}

Figure~\ref{fig:kernel_speedup} plots the speedup of the GPU runs over the CPU runs with 7 OpenMP threads for each computational kernel.
%Except for one kernel labeled \texttt{RiemannSolverInput}, speedups of more than 10X are achieved.
Speedups of more than 10X-40X are achieved.
%The non-performing \texttt{RiemannSolverInput} therefore warrants further investigation, which we have not done at this point.

Are we getting the speedups we can reasonably expect from offloading our kernels to the GPU via OpenMP \texttt{target} directives? 
To answer that question, we note the following facts.
%For this example fluid dynamics problem, most of the kernel algorithm are memory-bandwidth bound, since most of the work involve vector-like operations across all cells.
In this example fluid dynamics problem most of the kernels are memory-bandwidth bound, since most of the work involves vector-like operations across all cells.
Therefore we can at least expect performance improvements proportional to the ratio of the bandwidths of the GPU's HBM2 memory to the CPU's DDR4 main memory.
Summit's HBM2 theoretical peak bandwidth is $900$ GB/s,\footnote{https://www.olcf.ornl.gov/for-users/system-user-guides/summit/nvidia-v100-gpus/} while its DDR4 bandwidth is $135$ GB/s,\footnote{https://www.olcf.ornl.gov/for-users/system-user-guides/summit/system-overview/}
%The HBM2 to DDR4 bandwidth ratio is $\sim6.66\mathrm{X}$.
implying a HBM2 to DDR4 bandwidth ratio of $\sim6.66\mathrm{X}$.
%Since most of GPU offloaded kernels get speedup exceeding this ratio, it seems that we also get the extra benefit from the much higher computational power of the GPUs.
Since most of the GPU offloaded kernels get speedups exceeding this ratio, it seems that we also benefit from the much higher computational power of the GPUs.

\begin{figure}
\centering
\includegraphics[width=0.46\textwidth]{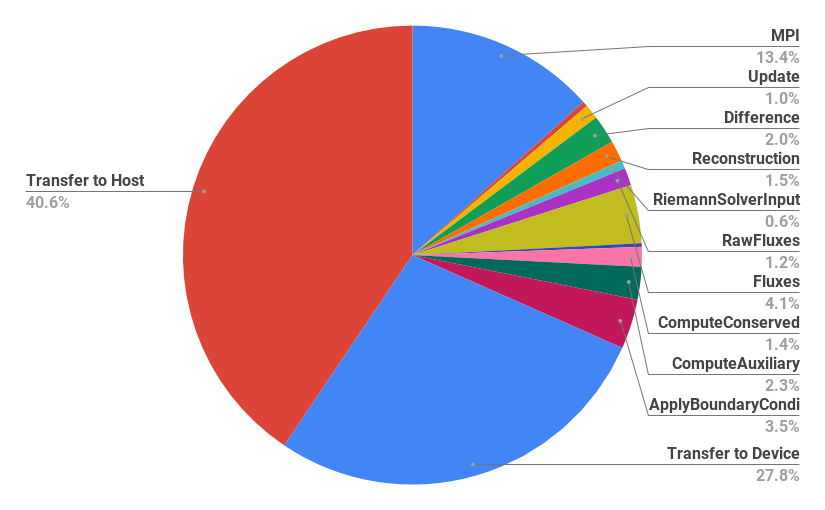}
\caption{Timing distribution of the different computational kernels, MPI communication, and data transfers in \texttt{RiemannProblem} when run with GPU.
More than 50\% of the time is due to data transfers between the host CPU and the GPU.} 
\label{fig:kernel_dist}
\end{figure}

%By looking at timing distributions of the different portions of the code, we can get a good idea of further low-hanging fruits for optimization.
%Figure \ref{fig:kernel_dist} shows the timing proportions of the different portions of the GPU version of 3D \texttt{RiemannProblem}.
Figure~\ref{fig:kernel_dist} shows the timing proportions of the different portions of the GPU runs of the 3D \texttt{RiemannProblem}, revealing prime targets for further attention.
%It is easy to see that data transfers between the GPU and host memory make up the largest portions of wall time for our application.
Despite overall speedups of 6X and beyond, the data transfers between the GPU and host memory loom large in taking more wall time than anything else in our application.
The primary cause of these data transfers is the need to move the fluid data for MPI communication (ghost exchanges and time step reduction) for every time step, since our MPI communication is done on the host CPU.
In the next section we discuss how we have alleviated these costs.

%\subsection{Sharing GPU with Multi-Process Server}

% \begin{itemize}
%  \item Plot overall speedup
%  \item Plot per kernel speedup for 1GPU, 6GPU, Many GPU
% \end{itemize}

\section{Pinned Memory}
\label{sec:BeyondOpenMP}

Figure \ref{fig:kernel_dist} suggests that further speedups can be achieved if we can reduce the time required for data transfers between host and device.
Host data that are allocated on a page-locked memory---more commonly known as \textit{pinned memory}---can be transferred between host and device much more efficiently than data allocated on pageable host memory.
However, OpenMP 4.5 does not provide a mechanism to allocate host data on pinned memory.
To do this we provide new Fortran routines \texttt{AllocateHost()} and \texttt{DeallocateHost()} as wrappers to the CUDA functions \texttt{cudaHostAlloc()} and \texttt{cudaFreeHost()}, respectively. 
We added an option to the initialization method of \texttt{StorageForm} to allocate the data member as pinned memory by calling the routine \texttt{AllocateHost()}, instead of using the the intrinsic Fortran statement \texttt{allocate}.

\begin{figure}
\centering
\includegraphics[width=0.46\textwidth]{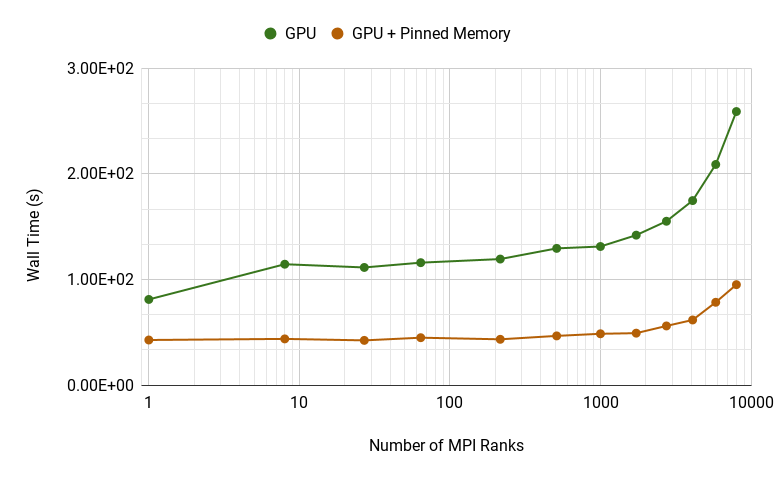}
\caption{Weak scaling of the 3D \texttt{RiemannProblem} for the GPU versions with and without pinned memory. Figure~\ref{fig:3D_WeakScaling}. 
The line labeled ``GPU + Pinned Memory'' shows the timings when \texttt{FluidCurrent} in Algorithm~\ref{alg:EvolveAccelerator} is allocated on pinned memory. 
Speedups of $\sim1.7 - 2.0 \mathrm{X}$ when pinned memory is used, resulting overall speedups of $\sim 12\mathrm{X}$ from 7 CPU threads.}
\label{fig:3D_WeakScaling_Pinned}
\end{figure}

\begin{figure}
\centering
\includegraphics[width=0.46\textwidth]{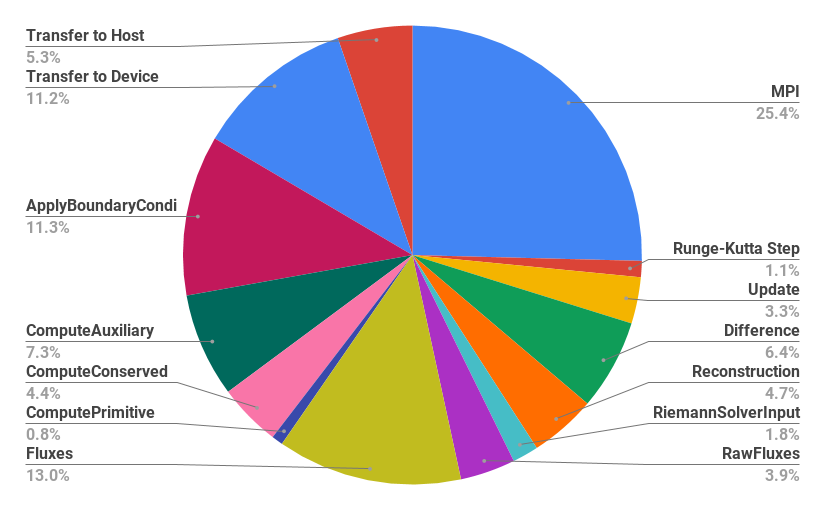}
\caption{Similar to Figure \ref{fig:kernel_dist} but with the use of pinned memory to optimize data transfer.
The data transfer time is reduced to less than 17\% of overall runtime.} 
\label{fig:kernel_dist_pinned}
\end{figure}

Figure \ref{fig:3D_WeakScaling_Pinned} shows the weak scaling plot of the 3D \texttt{RiemannProblem} of the GPU runs with and without the use of pinned memory. 
In this figure, the line labeled ``GPU'' is the same as the one in Figure~\ref{fig:3D_WeakScaling}. 
The line labeled ``GPU + Pinned Memory'' shows the timings when \texttt{FluidCurrent} in Algorithm~\ref{alg:EvolveAccelerator} is allocated on pinned memory.
We observe speedups between $\sim1.7 - 2.0 \mathrm{X}$ when pinned memory is used, resulting overall speedups of $\sim 12\mathrm{X}$ relative to 7 CPU threads. 
We plot the timing distribution of the different computational kernels when pinned memory is used in Figure \ref{fig:kernel_dist_pinned}.

\section{Conclusion}
\label{sec:Conclusion}

%On this paper we have described our efforts in using OpenMP to implement GPU-related infrastructure in \genasis\ \texttt{Basics} and port fluid dynamics kernels in the \texttt{RiemannProblem} example.
In this paper we describe our use of OpenMP 4.5 to implement GPU-related infrastructure in \genasis\ \texttt{Basics} and port fluid dynamics kernels in the \texttt{RiemannProblem} example.
%In our experience, these efforts have been relatively simple and straightforward, yielding very effective results with speedups of $\sim 4\mathrm{X}$ for the \texttt{RiemannProblem}. 
This has proved to be simple and effective, yielding speedups of $\sim 6\mathrm{X}$ for our \texttt{RiemannProblem}. 
In section \ref{sec:BeyondOpenMP} we went beyond the current OpenMP 4.5 specification to achieve even further speedups to $\sim 12\mathrm{X}$ using pinned memory.
These results further motivate the need for more advanced memory management with OpenMP, which is available in the new OpenMP 5.0 specification.

\section*{Acknowledgements}
This material is based upon work supported by the U.S. Department of Energy, Office of Science, Office of Nuclear Physics under contract number DE-AC05-00OR22725; the National Science Foundation under Grant No. 1535130.
This research used resources of the Oak Ridge Leadership Computing Facility, which is a DOE Office of Science User Facility supported under Contract DE-AC05-00OR22725.

\bibliographystyle{unsrt}
\bibliography{references}

\end{document}